\documentclass[a4paper, 11pt]{article}

\usepackage[utf8x]{inputenc}
\usepackage{epsfig,amssymb,euscript,xspace, color}
\usepackage{jheppub}
\usepackage[T1]{fontenc}
\usepackage{color}
\usepackage{amssymb}
\usepackage{amsfonts,amsmath,bm}
\usepackage{graphicx}
\usepackage{mathtools}
\usepackage{ragged2e}

\usepackage{dcolumn}
\usepackage{graphicx}
\usepackage{braket}
\usepackage{verbatim}
\usepackage[english]{babel}
\usepackage{hyperref}
\hypersetup{
citecolor=red,
colorlinks=true,
filecolor=red,
linkcolor=blue,
linktocpage=true,
urlcolor=blue
}
\usepackage[titletoc,toc,title]{appendix}
\numberwithin{equation}{section}
\usepackage{cleveref}
\usepackage{epsfig}
\usepackage{float}
\usepackage{amsfonts}
\usepackage{enumitem}
\usepackage[font={footnotesize,it}]{caption}

\newcommand{\be}{\begin{equation}}
\newcommand{\ee}{\end{equation}}
\newcommand{\eq}[1]{(\ref{#1})}
\newcommand\trick[1]{}
\newcommand{\bit}{\begin{itemize}}  \newcommand{\eit}{\end{itemize}}
\newcommand{\ben}{\begin{enumerate}}  \newcommand{\een}{\end{enumerate}}

\newcommand{\rf}[1]{(\ref{#1})}

\def\bd{\begin{document}}
\def\ed{\end{document}}
\def\bea{\begin{eqnarray}}
\def\eea{\end{eqnarray}}
\let\bm=\bibitem

\def\la{\langle}
\def\ra{\rangle}

\def\npb#1#2#3{Nucl. Phys. {\bf{B#1}} #3 (#2)}
\def\plb#1#2#3{Phys. Lett. {\bf{#1B}} #3 (#2)}
\def\prl#1#2#3{Phys. Rev. Lett. {\bf{#1}} #3 (#2)}
\def\prd#1#2#3{Phys. Rev. {D bf{#1}} #3 (#2)}
\def\cmp#1#2#3{Comm. Math. Phys. {\bf{#1}} #3 (#2)}
\def\cqg#1#2#3{Class. Quantum Grav. {\bf{#1}} #3 (#2)}
\def\nppsa#1#2#3{Nucl. Phys. B (Proc. Suppl.) {\bf{#1A}}#3 (#2)}
\def\ap#1#2#3{Ann. of Phys. {\bf{#1}} #3 (#2)}
\def\ijmp#1#2#3{Int. J. Mod. Phys. {\bf{A#1}} #3 (#2)}
\def\rmp#1#2#3{Rev. Mod. Phys. {\bf{#1}} #3 (#2)}
\def\mpla#1#2#3{Mod. Phys. Lett. {\bf A#1} #3 (#2)}
\def\jhep#1#2#3{J. High Energy Phys. {\bf #1} #3 (#2)}
\def\atmp#1#2#3{Adv. Theor. Math. Phys. {\bf #1} #3 (#2)}

\def\sst{\scriptscriptstyle}
\def\thetabar{\bar\theta}
\def\Tr{{\rm Tr}}
\def\one{\mbox{1 \kern-.59em {\rm l}}}

\def\a{\alpha}      \def\da{{\dot\alpha}}  \def\dA{{\dot A}}
\def\b{\beta}       \def\db{{\dot\beta}}
\def\g{\gamma}  \def\G{\Gamma}  \def\dc{{\dot\gamma}}
\def\d{\delta}  \def\D{\Delta}  \def\ddt{\dot\delta}
\def\e{\epsilon}
\def\ve{\varepsilon}
\def\uve{\upvarepsilon}
\def\f{\phi}    \def\F{\Phi}    \def\vvf{\f}
\def\vphi{\varphi}
\def\h{\eta}
\def\k{\kappa}
\def\l{\lambda} \def\L{\Lambda}
\def\m{\mu} \def\n{\nu}
\def\o{\omega}
\def\p{\pi} \def\P{\Pi}
\def\r{\rho}
\def\s{\sigma}  \def\S{\Sigma}
\def\t{\tau}
\def\th{\theta} \def\Th{\Theta} \def\vth{\vartheta}
\def\X{\Xeta}
\def\z{\zeta}

\def\na{\nabla}

\def\cA{{\cal A}} \def\cB{{\cal B}} \def\cC{{\cal C}}
\def\cD{{\cal D}} \def\cE{{\cal E}} \def\cF{{\cal F}}
\def\cG{{\cal G}} \def\cH{{\cal H}} \def\cI{{\cal I}}
\def\cJ{{\mathscr J}} \def\cK{{\cal K}} \def\cL{{\cal L}}
\def\cM{{\cal M}} \def\cN{{\cal N}} \def\cO{{\cal O}}
\def\cP{{\cal P}} \def\cQ{{\cal Q}} \def\cR{{\cal R}}
\def\cS{{\cal S}} \def\cT{{\cal T}} \def\cU{{\cal U}}
\def\cV{{\cal V}} \def\cW{{\cal W}} \def\cX{{\cal X}}
\def\cY{{\cal Y}} \def\cZ{{\cal Z}}
\def\ct{{\cal t}} \def\cw{{\mathcal{w}}} 

\def\ua{\underline{\alpha}}
\def\uc{\underline{\phantom{\alpha}}\!\!\!\gamma}
\def\um{\underline{\mu}}
\def\ud{\underline\delta}
\def\ue{\underline\epsilon}
\def\una{\underline a}\def\unA{\underline A}
\def\unb{\underline b}\def\unB{\underline B}
\def\unc{\underline c}\def\unC{\underline C}
\def\und{\underline d}\def\unD{\underline D}
\def\une{\underline e}\def\unE{\underline E}
\def\unf{\underline{\phantom{e}}\!\!\!\! f}\def\unF{\underline F}
\def\unm{\underline m}\def\unM{{\underline M}}
\def\unn{\underline n}\def\unN{{\underline N}}
\def\unp{\underline{\phantom{a}}\!\!\! p}\def\unP{\underline P}
\def\unq{\underline{\phantom{a}}\!\!\! q}
\def\unQ{\underline{\phantom{A}}\!\!\!\! Q}
\def\unH{\underline{H}}

\def\As {{A \hspace{-6.4pt} \slash}\;}
\def\bs {{b \hspace{-6.4pt} \slash}\;}
\def\Ds {{D \hspace{-6.4pt} \slash}\;}
\def\Gts {{\Gt \hspace{-6.4pt} \slash}\;}
\def\ds {{\del \hspace{-6.4pt} \slash}\;}
\def\ss {{\s \hspace{-6.4pt} \slash}\;}
\def\ks {{ k \hspace{-6.4pt} \slash}\;}
\def\ps {{p \hspace{-6.4pt} \slash}\;}
\def\xs {{x \hspace{-6.4pt} \slash}\;}
\def\pas {{{p_1} \hspace{-6.4pt} \slash}\;}
\def\pbs {{{p_2} \hspace{-6.4pt} \slash}\;}
\def\cFs {{{\cal F} \hspace{-6.4pt} \slash}\;}
\def\Dss {{D \hspace{-7.5pt} \slash}\;}
\def\dss {{\del \hspace{-7.0pt} \slash}\;}

\def\Ah{{\hat{A}}}
\def\Dh{{\hat{D}}}
\def\Gh{{\hat{G}}}
\def\Fh{{\hat{F}}}
\def\Ih{{\hat{I}}}
\def\Jh{{\hat{J}}}
\def\Kh{{\hat{K}}}
\def\Lh{{\hat{L}}}
\def\Ph{{\hat{P}}}
\def\Rh{{\hat{R}}}
\def\Vh{{\hat{V}}}
\def\Xh{{\hat{X}}}

\def\ah{{\hat{\a}}}
\def\bh{{\hat{\b}}}
\def\gh{{\hat{\g}}}
\def\dh{{\hat{\d}}}
\def\rh{{\hat{\r}}}
\def\hh{\hat{h}}
\def\uh{\hat{u}}
\def\xh{\hat{x}}
\def\yh{\hat{y}}
\def\ph{\hat{p}}
\def\xih{\hat{\xi}}
\def\chih{\hat{\chi}}
\def\Psih{\hat{\Psi}}
\def\phih{\hat{\phi}}

\def\psit{\tilde{\psi}}
\def\Psit{\tilde{\Psi}}
\def\Psibt{\tilde{\bar{Psi}}}

\def\lambdat{\tilde {\lambda}}
\def\st{\tilde{\sigma}}

\def\delt{\tilde{\delta}}
\def\Phit{\tilde{\Phi}}
\def\Phitb{\overline{\tilde{Phi}}}
\def\tht{\tilde{\th}}
\def\lt{\tilde{\l}}
\def\chit{\tilde{\chi}}
\def\phit{\tilde{\phi}}

\def\At{\tilde{A}}
\def\Bt{\tilde{B}}
\def\Ct{\tilde{C}}
\def\Dt{\tilde{D}}
\def\Et{\tilde{E}}
\def\Ft{\tilde{F}}
\def\Gt{\tilde{G}}
\def\Ht{\tilde{H}}
\def\It{\tilde{I}}
\def\Jt{\tilde{J}}
\def\Qt{\tilde{Q}}
\def\Rt{\tilde{R}}
\def\Mt{\tilde{M }}
\def\Nt{\tilde{N}}
\def\St{\tilde{S}}
\def\Vt{\tilde{V}}
\def\Xt{\tilde{X}}
\def\at{\tilde{a}}
\def\dt{\tilde{d}}
\def\htt{\tilde{h}}
\def\ft{\tilde{f}}
\def\gt{\tilde{g}}
\def\pt{\tilde{p}}
\def\qt{\tilde{q}}
\def\vt{\tilde{v}}
\def\nt{\tilde{n}}
\def\ut{\tilde{u}}
\def\wt{\tilde{w}}
\def\zt{\tilde{z}}
\def\xt{\tilde{x}}
\def\yt{\tilde{y}}
\def\Psit{\tilde{\Psi}}
\def\vphit{\tilde{\varphi}}
\def\tD{\tilde{\D}}

\def\eb{\bar{\epsilon}}
\def\delb{\bar{\partial}}
\def\thb{\bar{\theta}}
\def\mub{\bar{\mu}}
\def\lamb{\bar{\l}}
\def\psib{\bar{\psi}}
\def\sb{\bar{\sigma}}
\def\xib{\bar{\xi}}
\def\chib{\bar{\chi}}

\def\Psib{\bar{\Psi}}
\def\Phib{\bar{\Phi}}
\def\Lamb{\bar{\Lambda}}
\def\Sb{{\overline \Sigma}}
\def\cb{\bar{c}}
\def\hb{\bar{h}}
\def\qb{\bar{q}}
\def\wb{\bar{w}}
\def\ub{\bar{u}}
\def\zb{{\bar{z}}}
\def\Hb{\bar{H}}
\def\Qb{{\bar Q}}
\def\Omegab{\overline{\Omega}}
\def\ob{\overline{\omega}}

\def\Ab{{\overline A}} \def\Bb{{\overline B}} \def\Cb{{\overline C}}
\def\Db{{\overline D}} \def\Eb{{\overline E}} \def\Fb{{\overline F}}
\def\Gb{{\overline G}}
\def\Ib{{\overline I}}
\def\Jb{{\overline J}} \def\Kb{{\overline K}} \def\Lb{{\overline L}}
\def\Mb{{\overline M}} \def\Nb{{\overline N}} \def\Ob{{\overline O}}
\def\Pb{{\overline P}}  \def\Rb{{\overline R}}
 \def\Tb{{\overline T}} \def\Ub{{\overline U}}
\def\Vb{{\overline V}} \def\Wb{{\overline W}} \def\Xb{{\overline X}}
\def\Yb{{\overline Y}} \def\Zb{{\overline Z}}

\def\fb{{\overline f}}
\def\gb{{\overline g}}
\def\nb{{\overline n}}
\def\mb{{\overline m}}
\def\lb{{\overline l}}
\def\yb{{\overline y}}

\def\ldel{{\overleftarrow{\del}}}
\def\rdel{{\overrightarrow{\del}}}
\def\ldeldel{{\overleftarrow{\del^2}}}
\def\rdeldel{{\overrightarrow{\del^2}}}
\def\ldelb{{\overleftarrow{\bar{\del}}}}
\def\rdelb{{\overrightarrow{\bar{\del}}}}

\def\ba{{\bf a}}
\def\bk{{\bf k}}
\def\bl{{\bf l}}
\def\bp{{\bf p}}
\def\bq{{\bf q}}
\def\br{{\bf r}}
\def\bt{{\bf t}}
\def\bu{{\bf u}}
\def\bv{{\bf v}}
\def\bx{{\bf x}}
\def\by{{\bf y}}
\def\bA{{\bf A}}
\def\bR{{\bf R}}
\def\bV{{\bf V}}

\def\bz{{\boldsymbol{\zeta}}}

\def\bone{{\bf 1}}

\def\va{{\vec a}}
\def\vk{{\vec k}}
\def\vp{{\vec p}}
\def\vq{{\vec q}}
\def\vx{{\vec x}}
\def\vy{{\vec y}}
\def\vu{{\vec u}}
\def\vv{{\vec v}}
\def \vH{{\vec H}}
\def \vg{{\vec g}}

\def\vs{{\vec \sigma}}
\def\vtau{{\vec \tau}}

\newcommand{\ov}[1]{\overrightarrow{#1}}

\def\frA{\mathfrak{A}}
\def\frB{\mathfrak{B}}
\def\frC{\mathfrak{C}}
\def\frD{\mathfrak{D}}
\def\frE{\mathfrak{E}}
\def\frF{\mathfrak{F}}
\def\frG{\mathfrak{G}}
\def\frH{\mathfrak{H}}
\def\frM{\mathfrak{M}}
\def\frN{\mathfrak{N}}
\def\frR{\mathfrak{R}}
\def\frW{\mathfrak{W}}

\def\fra{\mathfrak{a}}
\def\frb{\mathfrak{b}}
\def\frf{\mathfrak{f}}
\def\frg{\mathfrak{g}}
\def\frh{\mathfrak{h}}
\def\frl{\mathfrak{l}}
\def\frs{\mathfrak{s}}
\def\fri{\mathfrak{i}}
\def\frj{\mathfrak{j}}

\def\ma{\mathfrak{a}}
\def\mg{\mathfrak{g}}
\def\mh{\mathfrak{h}}
\def\mR{\mathfrak{R}}
\def\mN{\mathfrak{N}}

\def\d{\delta}\def\D{\Delta}\def\ddt{\dot\delta}

\def\pa{\partial} \def\del{\partial}
\def\xx{\times}
\def\uno{\mbox{1 \kern-.59em {\rm l}}}

\def\trp{^{\top}}
\def\inv{^{-1}}
\def\dag{\dagger}
\def\pr{^{\prime}}

\def\rar{\rightarrow}
\def\lar{\leftarrow}
\def\lrar{\leftrightarrow}

\newcommand{\0}{\,\!}      
\def\one{1\!\!1\,\,}
\def\im{\imath}
\def\jm{\jmath}

\newcommand{\tr}{\mbox{tr}}
\newcommand{\slsh}[1]{/ \!\!\!\! #1}

\newcommand{\1}{\mbox{1}\hspace{-0.25em}\mbox{l}}

\def\vac{|0\rangle}
\def\lvac{\langle 0|}

\def\hlf{\frac{1}{2}}
\def\ove#1{\frac{1}{#1}}
\newcommand{\hot}[1]{\frac{#1}{2}}

\def\Box{\square}
\def\CC {\mathbb{C}}
\def\FF {\mathbb{F}}
\def\RR{\mathbb{R}}
\def\NN{\mathbb{N}}
\def\ZZ{\mathbb{Z}}
\def\bb#1{{\bf #1}}
\def\bcomment#1{}
\def\bfhat#1{{\bf \hat{#1}}}
\def\VEV#1{\left\langle #1\right\rangle}

\newcommand{\ex}[1]{{\rm e}^{#1}} \def\ii{{\rm i}}

\newcommand{\lrbrk}[1]{\left(#1\right)}
\newcommand{\lrsbrk}[1]{\left[#1\right]}
\newcommand{\sfrac}[2]{{\textstyle\frac{#1}{#2}}}

\def\stw{{\sqrt{2}}}

\def\rf {{\rm f}}
\def\ri {{\rm i}}
\def\rj {{\rm j}}
\def\rn {{\rm n}}
\def\rk {{\rm k}}
\def\rl {{\rm l}}
\def\rr {{\rm r}}
\def\rs {{\scriptscriptstyle \rm S}}
\def\rt {{\scriptscriptstyle \rm T}}

\def\rQ {{\scriptscriptstyle \rm \cQ}}
\def\rR {{\scriptscriptstyle \rm \cR}}

\def\cQb{{\cal \Qb}}
\def\cRb{{\cal \Rb}}
\def\cWb{{\cal \Wb}}

\def\fd {{\rm N}}
\def\afd {{\overline{\rm N}}}

\def \II {I\hspace{-.1em}I\hspace{.1em}}
\def \IIA {\mbox{\II A\hspace{.2em}}}
\def \IIB {\mbox{\II B\hspace{.2em}}}
\def \gs {g^s}
\def \ls {\lambda^s}

\def \I {{\cal I}}
\def \qs {q\hspace{-.53em}/\hspace{.15em}}
\def \ks {k\hspace{-.53em}/\hspace{.15em}}
\def \YM {{\mbox{\tiny YM}}}
\def \gym {g_{\YM}}

\def \Lc {\L_c}
\def\IR{\relax{\rm I\kern-.18em R}}
\def \id {{\bf 1}}

\def\cci{\ell}
\def\ccj{\ell'}

\def\bbq{\pmb{q}}
\def\bom{\pmb{\o}}
\def\bJ{\pmb{J}}
\def\bM{\pmb{M}}
\def\bB{\pmb{B}}
\def\bn{\pmb{n}}
\def\bE{\pmb{E}}

\title{\Large \bf\boldmath Massless Lifshitz Field Theory for Arbitrary $z$}

\author[a,b]{Jaydeep Kumar Basak}
\author[c,d]{Adrita Chakraborty}
\author[c,d,e]{Chong-Sun Chu}
\author[a,b,e]{Dimitrios Giataganas}
\author[d,e]{Himanshu Parihar}

\affiliation[a]{
	 Department of Physics, National Sun Yat-Sen University, Kaohsiung 80424, Taiwan}
\affiliation[b]{
	 Center for Theoretical and Computational Physics, Kaohsiung 80424, Taiwan}
 \affiliation[c]{
	Department of Physics, National Tsing-Hua University,
 Hsinchu 30013, Taiwan}
\affiliation[d]{Center of Theory and Computation,
National Tsing-Hua University,
 Hsinchu 30013, Taiwan}
\affiliation[e]{Physics Division,
    National Center for Theoretical Sciences,
     Taipei 10617, Taiwan}

\emailAdd{jkb.hep@gmail.com}
\emailAdd{w19603@phys.nthu.edu.tw}
\emailAdd{cschu@phys.nthu.edu.tw}
\emailAdd{dimitrios.giataganas@mail.nsysu.edu.tw}
\emailAdd{himansp@phys.ncts.ntu.edu.tw}

\abstract{\noindent By using the notion of fractional derivatives, we
  introduce a class
  of massless Lifshitz scalar field theory in (1+1)-dimension with an
  arbitrary anisotropy index $z$. The Lifshitz scale invariant ground
  state of the theory is constructed explicitly and takes  the form
  of Rokhsar-Kivelson (RK). We show that there is a continuous family of
  ground states with degeneracy  parameterized by the choice of
  solution to the equation of motion of an auxiliary
  classical system. The quantum mechanical path integral establishes a 2d/1d
  correspondence with the equal time correlation functions of the
  Lifshitz scalar field theory.
  We study the entanglement properties of the Lifshitz theory for arbitrary
  $z$ using the path integral representation.
  The entanglement measures are expressed in terms of certain cross
  ratio functions we specify, and satisfy the $c$-function monotonicity
  theorems.  We also consider the holographic description
  of the Lifshitz theory. In order to match with the field theory result
  for the entanglement entropy, we propose a $z$-dependent radius scale for the
  Lifshitz background. This relation is consistent with the
  $z$-dependent scaling symmetry respected by the Lifshitz vacuum.
  Furthermore, the
  time-like entanglement entropy is determined using holography. Our result
  suggests that there should exist a fundamental definition of time-like
  entanglement other than employing analytic continuation as
  performed in relativistic field theory. }

\begin{document}
	
	\maketitle
	\flushbottom
	\pagebreak

	\definecolor{orange}{rgb}{1.0, 0.49, 0.0}

\section{Introduction }

Invariance under global scaling transformation
\be \label{gls}
t\rightarrow \lambda t, \quad x^i\rightarrow \lambda x^i, \quad \l >0,
\ee
plays important roles in
physics. Apart from describing the fixed point of renormalization
group (RG) flow and critical phenomena in field theory, scaling symmetry
also finds important applications in particle physics at very high
energies and in the microwave background in cosmology.
In relativistic field theory, it has been shown that scale
symmetry always gets enhanced to the full conformal symmetry. This is not so
in non-relativistic field theory. A particularly well-known generalization of
the standard scaling symmetry is the Lifshitz scaling symmetry
\begin{equation}
t\rightarrow \lambda^z t, \quad  x^i \rightarrow \lambda x^i ,\quad
   \l>  0,
	\label{lss}
\end{equation}
which is characterized by  the
dynamical exponent $z \in \RR$.
Lifshitz symmetry has found important applications in
many physical systems. For instance, the $z=2$ Lifshitz scalar field theory
in (2+1) dimensions \cite{PhysRevB.23.4615}
\begin{equation}
    \mathcal{L}=\int d^2xdt
    \left[\left(\partial_t\phi\right)^2-\kappa^2\left(\nabla^2\phi\right)^2
      \right],
\end{equation}
is known to
describe the critical point of the well-known Rokhsar-Kivelson Quantum dimer
model \cite{PhysRevLett.61.2376, Henley_2004}.
The $z=2$ Lifshitz field theory in (3+1)
dimensions has appeared in the ghost condensation modified gravity scenario
\cite{Arkani-Hamed:2003pdi}
and describes a fluid with a non-relativistic dispersion relation
\be \label{dispersion}
\o^2 = \beta^2 k^{2\, z},
\ee
with  $\beta$ being some dimensionful energy scale of the theory.
Note that for $z<1$, the dispersion relation is acausal due to the
existence of  superluminal modes \cite{Koroteev:2011zz} and the violation of
the null energy conditions of the holographic dual \cite{Hoyos:2010at}.
Therefore we will focus on $z \geq 1$ in this paper.
As the Lifshitz scaling \eq{lss} is defined for general $z$,
it is interesting to ask
how the Lifshitz scaling \eq{lss} symmetry for other values of $z$
can be realized field theoretically.
In the literature, only Lifshitz scalar field theory with integer $z$
has been considered.
In this paper, we employ the mathematical notion of fractional
derivatives to propose an action for the
massless Lifshitz theory for arbitrary values of
$z$ in any dimensions.

Given a field theory, one of the first properties to understand is
its ground state, which is an important first step  to
the understanding of the physical system.
For example, in particle physics,
the  symmetry of the ground state dedicates
the particle spectrum and their interaction in the low energy theory.
In statistical mechanics, the ground state of a system provides most of the
 thermodynamic properties of the system. The
 entanglement properties of ground state
 has found intimate relation with topological order
 \cite{Hamma:2005hga,Kitaev:2005dm, PhysRevLett.96.110405} and
 quantum critical phenomena \cite{Osterloh2002608, Osborne:2002zz,
   Vidal:2002rm, Calabrese:2004eu} in quantum many body system.
In the literature, it is known that the $z=2$ massless Lifshitz theory has a
ground state  which takes on a special form of Rokhsar-Kivelson (RK)
\cite{PhysRevLett.61.2376},
in the sense that the ground state is local and is given by a superposition of
quantum states with a quantum mechanical amplitude  $c[\phi]$ specified
by the probability distribution of a classical system.
Once we have constructed the massless Lifshitz theory for arbitrary
$z$, it is interesting to construct its ground state and
examine  whether the ground state is still of the form of RK,
or is it just for the special value of $z=2$. The point $z=2$ is in fact
somewhat special
since it is known that the Lifshitz scaling transformation \eq{lss}
combines with the Galilean boost to the Schrodinger group only for this value.
This is analogous
to the situation in relativistic theory ($z=1$)
where the scaling symmetry \eq{gls}
combines with the Lorentz boosts to the conformal group. Therefore, it is
meaningful to ask if the RK form of the ground state is
a consequence of the Schrodinger symmetry or not.
In this paper, we show that the Lifshitz
ground state of the (1+1)-dimensional
massless theory always take the form of RK, even when there
is no presence of Schrodinger symmetry. It turns out that except for $z=2$,
there is a continuous family of degenerate Lifshitz ground states for $z>1$.
We show how the correlators in these Lifshitz ground state
can be computed in terms of the path integral of a
1-dimensional auxiliary quantum mechanical system.

In a relativistic quantum field theory, the
Reeh-Schlieder theorem \cite{Reeh:1961ujh} states that
all field variables in any one region
are entangled with variables in other regions. This means
the entanglement is  so strong that it leads to
an ultraviolet divergent entanglement entropy between adjoining
regions in  spacetime. In the article \cite{Witten:2018zxz}, it is explained how
this ultraviolet divergence of entanglement is in fact
a property of  the algebras of observables. As entanglement has not much to do
with the speed of light, we expect to have a similar story for the
non-relativistic Lifshitz field theory. However, the detailed form of
entanglement encoded in different entanglement observables will
certainly be different.
Previously, entanglement entropy in anisotropic Lifshitz field
theories have been analyzed in a series of literature in the context
of the quantum Lifshitz model (QLM) \cite {Ardonne_2004} which is a
special type of Lifshitz field theory in $(2+1)$-dimensions with $z=2$
\cite{Fradkin:2006mb,Fradkin:2009dus, PhysRevB.80.184421,Hsu:2008af,Hsu:2010ag,
  Oshikawa:2010kv,
  PhysRevLett.107.020402,Zhou:2016ykv,MohammadiMozaffar:2017nri,
  Berthiere:2019lks,Boudreault:2021pgj}. Subsequently
the QLM was generalized to $(d+1)$-dimensions with $z=d$, $d$ being the
number of spatial dimensions. The holographic realization of
the Lifshitz symmetry for the arbitrary $z$ was studied in
\cite{Hosseini:2015gua} using the RT prescription in higher curvature
Lifshitz gravity. Further recent progress has been made in the
holographic interpretation of such theory by computing both
equal-space and equal-time two-point correlation functions and their
exact matching in both field theory and bulk sides \cite{
  Ker_nen_2012, Keranen:2016ija, Park:2022mxj}. The entanglement
entropy has been studied for the $z=2$ quantum Lifshitz models
\cite{Angel-Ramelli:2019nji} followed by the investigation of
logarithmic negativity in $(1+1)$ and
$(2+1)$-dimensions \cite{MohammadiMozaffar:2017chk,Angel-Ramelli:2020wfo}.
Added to this,
entanglement entropy of Lifshitz field theory for arbitrary $z$
has also been investigated with cMERA techniques
\cite{Nozaki:2012zj,He:2017wla,Gentle:2017ywk}.  Very recently,
reflected entropy \cite{Dutta:2019gen} and Markov gap \cite{Hayden:2021gno}
were analyzed
for (1+1)-dimensional Lifshitz field theory with $z=2$ in
\cite{Berthiere:2023bwn}.

Once we have found the massless Lifshitz field theory and its ground
states for arbitrary $z$, it is an interesting question to study how
the entanglement properties of the Lifshitz theory depend on the time
anisotropy index $z$. In this paper, we analyze the entanglement and
correlation properties of the (1+1)-dimensional Lifshitz theory for a
number of observables.  We find that the entanglement entropy
increases with respect to $z$. This is in agreement with the
expectations that the increase of $z$ leads to long-range interaction
of the theory which typically translates to the growing of the
entanglement entropy. Using the kernel of the theory, we extend our
computations to the system of two adjacent and disjoint intervals to
analytically obtain the R\'{e}nyi entropy, reflected entropy and the
Markov gap. It is worthy to note that our measures can be expressed in
terms of cross ratios with components of the form $\eta(l,z)\sim
(\prod_i l_i^{z-1})/(\sum_j l_j^{z-1})$ that depend on the lengths on
the intervals $l_i$. Moreover, the standard $c$-function monotonicity
defined via the entanglement entropy along the RG flow, is satisfied
and its satisfaction turns out to be equivalent to the satisfaction of
the null energy conditions and the presence of causality in the theory
\cite{Chu:2019uoh}.

The holographic study of our class of massless Lifshitz scalar theory is
also interesting. It has been proposed \cite{Kachru:2008yh} that  the
holographic dual of a general Lifshitz field theory in $(d+1)$-dimensions
is given by the bulk metric
\begin{equation}
  ds^2=L^2\left[-\frac{dt^2}{r^{2z}} +\frac{d\vx^2}{r^2} +\frac{dr^2}{r^2}
    \right].
\end{equation}
The metric is not Lorentz-invariant,
but supports the scaling symmetry given by
\begin{equation}
  t\rightarrow\lambda^{z}t,~~x\rightarrow \lambda x,~~r\rightarrow
  \lambda r,
        \label{chosen scaling}
\end{equation}
that is consistent with the Lifshitz symmetry \eq{lss}
of the field theory. Such bulk solutions can be understood as arising
from the Einstein-Proca type bulk action which consists of a massive
vector field \cite{Taylor:2008tg,Taylor:2015glc}.
Using the RT formula, we compute the entanglement  entropy for the
strongly coupled holographic Lifshitz field theory.
As we will show in section 4, the form of the holographic entropy
agrees with the result \eq{ee-inf} of the free theory. This is
because the form of the UV divergent part of entanglement entropy
is universal. It is possible that, just
as in the case of 2d conformal field theory,
the coefficient of the entanglement entropy is protected by a
non-renormalization theorem. Assuming so, we are leaded to propose a
relation between the radius of curvature $L$ of the Lifshitz
background and the Newton constant.  This is analogous to the
Brown-Henneaux relation \cite{Brown:1986nw} $\frac{3R}{2l_P} =c$ for
the standard AdS${}_3$/CFT${}_2$ duality.  The relation has a novel
$z$-dependence and is consistent with the fact that the Lifshitz
vacuum respects a $z$-dependent scaling symmetry.  Furthermore, the
anisotropy between the temporal and spatial directions in LFTs makes
it interesting to investigate the behavior of entanglement entropy in
the time direction.  In this context, we compute the time-like
entanglement entropy using the holographic
prescription of \cite{Doi:2022iyj,Doi:2023zaf} (See
\cite{Wang:2018jva, Narayan:2022afv, Li:2022tsv, Jiang:2023ffu,
  Narayan:2023ebn, Jiang:2023loq, Chu:2023zah, He:2023ubi,
  Narayan:2023zen, Grieninger:2023knz} for recent developments). The
holographic result is $z$-dependent and suggests that there exists a
fundamental definition of time-like entanglement entropy other than
employing analytic continuation.

\section{Massless Lifshitz scalar theory and ground state}

In this section, we employ the definition of fractional calculus to introduce
the Lifshitz scalar field theory for an arbitrary real value of $z$.
For  the case of $(1+1)$-dimensions,
we show that
the ground state of the massless
Lifshitz field theory can be
constructed explicitly and takes on the form of RK. We call these Lifshitz
ground states as they are Lifshitz scaling invariant.
We show  that the theory possess a 2d/1d duality in that
   the equal time correlation functions of the 2-dimensional
  Lifshitz theory can be determined in
  terms of the path integral of a 1-dimensional quantum mechanical system.
  We also show that the trace involving powers of the reduced density
  matrix can be computed explicitly without introducing the twist operators.

\subsection{Massless Lifshitz scalar field theory in $(d+1)$-dimensions}

We start by considering the following action for the massless Lifshitz
scalar field
  theory in $(d+1)$-dimensions with arbitrary critical exponent $z$,
  \footnote{
Another scalar field theory realization of the Lifshitz scaling symmetry
is given by the action
\be \label{az2}
S=\frac{1}{2}\int dt d^{d}x \left[\left(\partial^{1/z}_{t}\phi\right)^2
   -\kappa^2\left(\del_i \phi\right)^2\right].
\ee
This is an interesting theory as well. However unlike \eq{az},
the action \eq{az2} does not admit a canonical
formulation and the quantization of the model becomes obscure.
}
\begin{equation}
  S=\frac{1}{2}\int dt d^{d}x \left[\left(\partial_{t}\phi\right)^2
    -\kappa^2 \phi \left(-\del_i^2\right)^z \phi\right],
	\label{az}
\end{equation}
where  $\k>0$ is a constant and
the operator $(-\del_i^2)^z$ is diagonal
$(-\del_i^2)^z e^{ikx} = k^{2z} e^{ikx}$ on the plane wave basis.
This form of action has also been considered independently 
in \cite{Keranen:2016ija} for integer $z$
and in \cite{Benedetti:2023pbt} for arbitrary  $z$.
The above action is invariant under the Lifshitz scaling symmetry
\eq{lss} and the transformation
\be \label{D-phi}
\phi \to \l^{-\Delta_\phi}\phi, \quad \mbox{where}\quad
  \Delta_\phi = \frac{d-z}{2}.
  \ee
Here $\D_\phi$ is the scaling dimension of the scalar field $\phi$.
Note that a self-interacting potential $\phi^\frac{d+z}{\D_\phi}$ with
fractional power can be added to the theory \eq{az}
and still preserve the
Lifshitz scaling symmetry.
For $z=1$, the massless action \eq{az} is Lorentzian invariant,
which is extended to the full conformal symmetry due to the presence of
scaling symmetry.
For $z =2$, the theory has the Galilean boost, which is extended to the
Schrodinger symmetry due to the presence of the Lifshitz scaling symmetry.
For other values of $z$, the theory has the Lifshitz symmetry group which
does not admit the Galilean boost.

The Green's function of the theory
\be
(\partial_t^2 + \kappa^2(-\del_i^2)^z ) G(x,x') = -\d^{(d+1)}(x-x'),
\ee
which is given by
\be
G(x,x') = c \int dk_0 d^{d} \bk \; \frac{e^{ik\cdot x}}{k_0^2 - \k^2 \bk^{2z}},
\ee
where $k\cdot x = k_0 t - \bk \cdot \bx$ and
$c = 1/(2\pi)^{d+1}$ is a normalization constant.
For the case of equal-time and equal-space separations, $G$ can be
evaluated using the method of
residues. One can obtain the equal-time correlators as
\be \label{g1}
G(\bx, \bx') \sim \frac{1}{|\bx - \bx'|^{2\D_\phi}},
\ee
and the equal-space correlators given by
\be \label{g2}
G(t,t') \sim \frac{1}{|t - t'|^{2\D_\phi/z}}.
\ee
For general configuration of positions, the expression for
$G$ is quite complicated and is given by integrals of
hypergeometric functions.
Previously, the Green's function of the theory \eq{az}
was determined for general point configurations for
the case $z =d$ \cite{Keranen:2016ija}. In this case the
scalar field has vanishing scaling dimensions and the two point functions
display a logarithmic singularity at short distance.
Recently, the  results  \eq{g1} and \eq{g2} were also
obtained by \cite{Park:2022mxj}. Here, we consider
the massless Lifshitz theory \eq{az} defined for arbitrary $z$ and obtain
these results directly as  special cases.

\subsection{Massless Lifshitz scalar theory and Lifshitz ground state
  in $(1+1)$ dimensions}

In this paper, we will consider the Lifshitz theory
in (1+1)-dimensions for arbitrary $z > 1$.
As shown 
in the appendix A of \cite{Keranen:2016ija} where the analysis
actually applies for arbitrary value of $z$,
the theory \eq{az} admits a RK vacuum. However, in addition to
the RK form of the vacuum,
one is also interested in the properties of the vacuum, e.g.
expectation value \eq{vev} of operators or entanglement properties.
As we will see below, many interesting properties of the RK vacuum
requires the knowledge of the solution to the equation of motion of
an associated 1-dimensional quantum system. For the theory \eq{az},
the equation of motion takes the form $ \del^z_x \phi =0 $, where
$\del^z_x:= (-\del_x^2)^{z/2}$ is a fractional
derivative with the modulus  $|k|^z$ as its eigenvalue. This is a different
fractional derivative
from the one \eq{del-def} we are going to define below,
and the construction of solution appears to be
more nontrivial. 
Instead let us consider the
following definition of the Lifshitz theory
\be  \label{az3}
 S=\frac{1}{2}\int dt d x \left[\left(\partial_{t}\phi\right)^2
    -\kappa^2  (\nabla_x^z\phi)^2\right],
 \ee
where the fractional derivative $\nabla_x^z$ is defined as
 \be \label{del-def}
\nabla_x^z e^{ikx} := (ik)^z e^{ikx}
\ee
and a choice of the branch of the multi-valued function $w^z, w \in \CC$
is chosen. For example,
${\rm Log} w = {\rm Log} |w| + i \theta, -3\pi/2<\theta \leq \pi/2$ with a cut
at the positive imaginary axis.
Here, the fractional derivative is defined through \eq{del-def}
via some generalized notion of the Fourier analysis which involves some
appropriate momentum contour integral,
see, e.g., example in appendix \ref{Frac-deriv}.
We note that while the action \eq{az3}
is equivalent to \eq{az} when written in momentum
space, \eq{az3} and \eq{az}  are different in the presence of boundary.
The discussion of boundary Lifshitz theory is interesting,
and will be left for future consideration.

The construction of RK vacuum for  the theory \eq{az3}
resembles to that of \cite{Keranen:2016ija}.
The Hamiltonian of the theory
\begin{equation}
			H=\frac{1}{2}\int
                        dx\left(\Pi^2(x)+\kappa^2(\nabla_x^z\phi)^2\right),
\end{equation}
can be written in the factorized form
\begin{equation} \label{H1}
			H= \int dx A^{\dagger}(x) A(x)+E_{\text{vac}},
\end{equation}
where $E_{\rm vac} :=  \int \frac{dx dk}{2\pi} \frac{1}{2}(ik)^z$
is a real UV divergent constant
and
\begin{equation}\label{annihilation}
A(x)=\frac{1}{\sqrt{2}}\left( i\Pi(x) +\kappa\nabla_x^z{\phi}(x)\right),
\end{equation}
\begin{equation}	\label{creation}
A^{\dagger}(x)=\frac{1}{\sqrt{2}} \left(-i\Pi(x)+\kappa\nabla_x^z\phi(x)\right)
\end{equation}
are  generalized annihilation and creation
operators. They satisfy the following commutation relation
\be
[A(x), A^\dagger (x')] = \k \nabla_x^z \d(x-x')
\ee
where
$\left[\phi(x),\Pi(x')\right]=i\delta(x-x^{'})$ is used
for the canonical momentum $\Pi$.
Now since  $\int dx A^{\dagger}(x)A(x)$ is positive  semi-definite,
a ground state of the theory may be defined by using the
position space annihilation operator
\begin{equation} \label{AP}
			A(x)|\Psi_0\rangle=0, \quad \forall x.
\end{equation}
Note that
\be
A(x) \to \l^{z/2} A(x)
\ee
under the Lifshitz scaling \eq{lss}. Therefore $\ket{\Psi_0}$ is
Lifshitz scaling invariant and
we will call such $\ket{\Psi_0}$ a Lifshitz ground state.

In the Schrodinger representation $\Pi(x)=-i\frac{\partial}{\partial\phi(x)}$,
\eq{AP} turns into a differential equation for $|\Psi_0 \ra$
\begin{equation}	\label{functional equation}
\left[\frac{\delta}{\delta\phi}+\k\; \nabla^z_x\phi \right]
                  |\Psi_0\rangle=0 .
\end{equation}
This can be solved easily and the ground state of the Lifshitz theory is
given by
\begin{equation} \label{gsw}
  |\Psi_0\rangle=\frac{1}{\sqrt{\mathcal{Z}}} \int \cD \phi \;
  e^{-S_{\rm cl}[\phi]/2} |\phi\ra, \quad
   S_{\rm cl}[\phi] :=  \kappa
  \int \left(\nabla_x^{\frac{z}{2}}\phi\right)^2 dx ,
\end{equation}
where
\begin{equation}
\mathcal{Z}=\int \mathcal{D}\phi e^{-S_{\rm cl}[\phi]}
\end{equation}
is  a normalization factor. It is interesting to note
  that the ground state wavefunctional \eq{gsw} takes
  the form of RK \cite{PhysRevLett.61.2376}, it
  is given
  by a superposition of quantum states with a quantum mechanical amplitude
  $c[\phi]$
  given by the Boltzmann weight of a classical system, $c[\phi] \propto
  e^{-S_{\rm cl}[\phi]/2}$. In this case, the classical system is a 1d particle theory
  with action given by $S_{\rm cl} [\phi]/2$, and as such, $\cZ$ get the
  interpretation as the corresponding partition function.

With the vacuum \eq{gsw}, the expectation value of operators at equal time
 is given by the path integral
  \be \label{vev}
\bra{\Psi_0} O_1(\phi(x_1)) O_2(\phi(x_2)) \cdots \ket{\Psi_0}
= \frac{1}{\cZ} \int \cD \phi  e^{-S_{\rm cl}[\phi]}
O_1(\phi(x_1)) O_2(\phi(x_2)) \cdots.
\ee
Note that  path integral \eq{vev} is not  to be
confused with the ordinary path integral expression of the original
Lifshitz theory \eq{az}, which
involves the 2d Lifshitz action instead of the
action $S_{\rm cl}$ of the auxiliary quantum mechanical system.
This equivalence of the 2-dimensional field theory with
a 1-dimensional system here
applies only for the equal time correlation functions and
is due to the specific RK form of the Lifshitz vacuum.
While this relation is interesting, the path integral
\eq{vev} requires further specification before it can be computed explicitly.
To explain this, let us consider a path integral over $\phi$ satisfying
an arbitrary
specified condition of the form $\phi(x_i)= \phi_i,
\phi(x_f) =\phi_f$,
\be
I_n:= \int_{\phi(x_i) = \phi_i}^{\phi(x_f) =\phi_f}  \cD \phi\;
\Psi_0^n(\phi),
\ee
where $\Psi_0 (\phi) = \langle \phi | \Psi_0 \rangle = e^{-S_{\rm cl}[\phi]/2}$
is
the ground state wavefunction and $n$ is an arbitrary positive integer.
As usual,
the integral can be evaluated by integrating out the fluctuations around
the classical solution $\phi_c$ to the equation of motion
\be \label{eom-rk}
\nabla_x^z \phi =0
\ee
of $S_{\rm cl}$.
This gives \cite{Feynman:100771}
\be
I_n = e^{-n S_{\rm cl}[\phi_c]/2} F(x_i,x_f),
\ee
where
$F(x_i,x_f):=\int \cD \d \phi\;
e^{-nS_{\rm cl}(\d \phi)/2}$ is obtained by integrating out the
fluctuation that satisfies the
conditions $\d \phi(x_i) =\d \phi(x_f) = 0$ and is
a function of $x_i, x_f$ only.

Except for the case of $z=2$,
the condition 
\be \label{bc-rk}
\phi(x_i) = \phi_i, \quad \phi(x_f) =\phi_f
\ee
alone is
not enough to fix an unique solution to
\eq{eom-rk}
for general $z>2$. 
This is clear for integer $z >2$ as the equation has derivative higher
than
order 2. To construct the solution to \eq{eom-rk} for general
$z$, we can make use of the definition \eq{del-def} for the fractional derivative and use the Fourier analysis with appropriate choice of integral contour as described in appendix \ref{Frac-deriv}. Now using \eq{del-x},  the equation of motion
\eq{eom-rk}  is immediately solved by
functions 
\be
(x-a)^{z-n}, \quad x\geq a, \qquad \mbox{for $n =1,2, \cdots$},
\ee
for arbitrary $a$. 
The
general solution to \eq{eom-rk} and \eq{bc-rk} is then
given by
\be \label{soln1}
\phi = (\phi_f -\phi_i) \sum_{n=1}^{N_z}\frac{ c_n }{l^{z-n}}(x-x_i)^{z-n}
+ \phi_i,
\qquad \sum_n c_n =1,
\ee
where $l := x_f -x_i$ and there $c_n$ are arbitrary constants.
Note that an upper bound $N_f$ is needed in \eq{soln1}
in order for $\phi$ to be finite at the endpoint $x=x_i$. 
For non-integer
$z$, $N_z =[z]$ and there are $[z]-1$ free parameters.
For integer $z$, $N_z = z-1$ as the term $n=z$ is a
constant and this has been taken into account already in \eq{soln1}.
Therefore
in this case there are $z-2$ free parameters. This analysis
is consistent with the case of $z=2$.
We can make a change of variable to $t := (x-x_i)/l$, then
\be \label{soln-g}
\phi = \phi_i +  (\phi_f -\phi_i)g (t),
\ee
with $g(t)$ given by the ``polynomial''
\be \label{poly-g}
 g(t): = \sum_{n=1}^{[z]} c_n t^{z-n}, \quad \sum_n c_n =1.
 \ee
Note that $g(t)$ is independent of the boundary condition \eq{bc-rk}.
Therefore, the evaluation of the path integral
\eq{vev} requires a specification of the function $g$ only once and for
all.
We may interpret this as saying that the Lifshitz theory has a family of
ground states whose degeneracy is parameterized by the function $g$.

Back to the vacuum
expectation value \eq{vev}, we can decompose the functional integration
in regions and rewrite it as an integral of the
insertion points $x_i$ over the product of the $O_i$'s with the
path integral propagator
\be \label{K-def}
  K(\phi_i, \phi_f; x_i, x_f) = \int_{\phi(x_i)
= \phi_i}^{\phi(x_f) =\phi_f}
  \cD \phi
  \exp \left( - \int_{x_i}^{x_f}
\left(\nabla_x^{\frac{z}{2}}\phi\right)^2 dx \right).
  \ee
  With a choice of $g$ made for the theory, the kernel $K$ is well defined
 and one obtains
\be \label{K}
K(\phi_i, \phi_f; l) = \sqrt{\frac{\g}{\pi l^{z-1}}}
e^{-\g (\phi_f-\phi_i)^2/l^{z-1}},
\ee
where $\g := \k c$ and $c := \int_0^1 (\del_t^{z/2} g(t))^2 dt$
is a constant that depends on the set $ \{c_n \}$ of
parameters. Explicitly, it is
\be
c =  \sum_{n,m} \frac{c_n c_m}{z-n-m+1}
\frac{\G(z-n+1) \G(z-m+1)}{\G(z/2-n+1) \G(z/2-m+1)}.
\ee
For example, for the solution
\be
\phi = \frac{\phi_f - \phi_i}{l^{z-1}}(x-x_i)^{z-1} + \phi_i,
\ee
which satisfies the Neumann boundary conditions at $x=x_i$:
\be
\nabla_x^{z-n} \phi |_{x=x_i} =0, \quad n = 2, 3, \cdots, [z],
\ee
we have
\be
\g := \frac{\k}{z-1}
\left(\frac{\Gamma(z)}{\Gamma\left(\frac{z}{2}\right)}\right)^2.
\ee
Note that
for any given $g$, the kernel $K$ satisfies the same composition property,
\be
\int \cD
\phi' K(\phi_1, \phi'; l_1) K(\phi', \phi_2; l_2) = K(\phi_1, \phi_2; l_{12}),
\ee
where $l_{12}$ is given by
\be \label{harm}
\frac{1}{l_{12}^{z-1}} = \frac{1}{l_1^{z-1}} + \frac{1}{l_2^{z-1}}.
\ee
In the following, we will consider the entanglement properties for a general
Lifshitz ground state.
It turns out that
the choice of the ground state only affects the constant term in
the entanglement quantities we computed for the Lifshitz theory.

Finally, we remark that in addition to the Lifshitz ground state $\ket{\Psi_0}$
which is defined in terms of the  position space annihilation
operator $A(x)$, one may also
consider the Fock vacuum $\ket{0}$ defined by the momentum space
annihilation operator $a_k$ obtained from the  canonical quantization of
the theory. Although also a ground state,
$\ket{0}$ is different from $\ket{\Psi_0}$ since $A(x)$ is given
by a nontrivial Bogoliubov transformation of the  momentum space
operators $a_k, a_k^\dag$.
Our analysis for the Lifshitz ground state holds valid only for $z>1$.
For example,
the RK vacuum of Lifshitz theory is not
defined for $z=1$ while the Fock vacuum is defined for $z=1$
and is continuous there. We will see below that they display
completely different entanglement properties.

\subsection{Reduced density matrix and replica}

\begin{figure}[H]
			\centering
                        \includegraphics[scale=1.2]{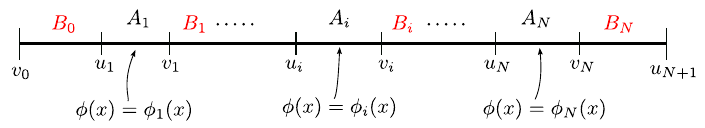}
			\caption{Schematic for the configuration consisting of intervals $A$ and $B$.}
			\label{generic}
  \end{figure}

We are interested in the entanglement properties of the Lifshitz ground state
$\ket{\Psi_0}$ of the theory. The  density matrix is given by
  \be
  \rho = \frac{1}{\cZ} \int \cD \phi \cD \phi' e^{-\frac{1}{2}(S_{\rm cl}[\phi]+
    S_{\rm cl}[\phi'])} \ket{\phi}\bra{\phi'}.
  \ee
 Consider a subsystem
  \be
  A := \bigcup_{i=1}^N A_i
  \ee
  as depicted in figure \ref{generic},
  which consists of $N$ intervals $A_i
  = (u_i, v_i)$, where $u_i <v_i$ are the endpoints of the interval $A_i$ and
  $i =1, \cdots, N$. Let us denote the complementary intervals as
  $B_i = (v_i, u_{i+1})$, $i =0 ,1, \cdots, N$ where $v_0$ (resp. $u_{N+1}$)
  denotes the
  coordinate of the left (resp. right) boundary of the total system.

  The reduced density matrix $\r_A = \tr_B \r$ is obtained by tracing over
  $B := \bigcup_{i=0}^N B_i$, the complement of $A$. Explicitly, it has the
  matrix elements
  \be
  (\r_A)_{\phi_A', \phi_A''} =  \frac{1}{\cZ} \int \cD \phi_B
  (\bra{\phi_B}\bra{\phi_A'}) \ket{\Psi_0}
  \bra{\Psi_0} (\ket{\phi_A''} \ket{\phi_B}).
  \ee
  Here $\phi_A',  \phi_A''$ are specified by their values over $A$:
    \be
    \phi_A'(x) = \{\phi_i'(x)\; \mbox{for $x \in A_i$, $i =1, \cdots, N$}\},
    \qquad
  \phi_A''(x) = \{\phi_i''(x)\; \mbox{for $x \in A_i$, $i =1, \cdots, N$}\}.
  \ee
  Evaluating the path integral, we obtain
  \be
  (\r_A)_{\phi_A', \phi_A''} = \frac{1}{\cZ} e^{-\frac{1}{2} S_{\rm cl}[\phi_A']
    -\frac{1}{2} S_{\rm cl}[\phi_A'']} \prod_{i=1}^N K(v_i, u_{i+1}),
  \ee
  where $K(v_i, u_{i+1})$ is the propagator \eq{K-def}  with the corresponding
  boundary values of $\phi$
  specified by $\b_i:= \phi_i(v_i)$ and $\a_{i+1}:=\phi_{i+1}(u_{i+1})$.
  We have suppressed the appearance of the boundary values of $\phi$ in the
  notation here.
  As a result, the trace
  $\cZ_1 := \int \cD \phi_A (\r_A)_{\phi_A,\phi_A}$
  is given by
  \be \label{Z1}
  \cZ_1 =\frac{1}{\cZ}
  \int_{-\infty}^\infty d\a_1d\b_1 \cdots d\a_N d\b_N
  \prod_{i=1}^N K(u_i,v_i) K(v_i,u_{i+1}).
  \ee
  Similarly the $n$-th power of the reduced density can be computed
  \be
  (\r_A^n)_{\phi_A', \phi_A''} =\frac{1}{\cZ^n}  e^{-\frac{1}{2} S_{\rm cl}[\phi_A']
    -\frac{1}{2} S_{\rm cl}[\phi_A'']}
  (\prod_{i=1}^N K(u_i, v_i))^{n-1} (\prod_{i=1}^N K(v_i, u_{i+1}))^n
  \ee
  and
the trace
  $\cZ_n := \int \cD \phi_A (\r_A^n)_{\phi_A,\phi_A}$
  is given by
  \be \label{Zn}
  \cZ_n  = \frac{1}{\cZ^n}
  \int_{-\infty}^\infty d\a_1d\b_1 \cdots d\a_N d\b_N
  \prod_{i=1}^N K^n(u_i,v_i)  \prod_{i=1}^N K^n(v_i,u_{i+1}).
  \ee
  Note that  we have assumed Dirichlet boundary condition in the above and so
  there is no  integration over the fields  $\b_0$ and $\a_{N+1}$
  at the boundary of the theory. Such an integration would be needed if free
  boundary is considered.
  We remark that
  the trace $\cZ_n$ of the density matrix
  has an interpretation as a partition
  function $\cZ_n = \int_{M_n} \cD \phi e^{-S_{\rm cl}}$
  over a $n$-branched covered manifold $M_n$. In CFT,
  the partition function can be readily computed  with
  the use of twisted operators. However, it is not clear how to introduce
  the twist operator in non-conformal field theory. Instead,
  we find that $\cZ_n$ can be determined directly in the Lifshitz
  theory  without using the formalism of twist operators
  due to the specific RK form of the
  Lifshitz vacuum.

\section{Entanglement}
  \label{EE-Lifshitz field theory-z}

Quantum information theory has found many
applications in our understanding of QFT, condensed matter physics,
quantum gravity and so on with a central aim to measure the degree of
correlations between different subsystems of a quantum system. Amongst
various quantum entanglement measures, entanglement entropy is the
most properly understood entanglement measure in different aspects
from QFTs to holography.  The entanglement entropy associated with a
subsystem $A$ when the full bipartite system $A\cup B$ is specified by a
density matrix $\rho$ with total Hilbert space
$\mathcal{H}=\mathcal{H}_A\otimes \mathcal{H}_B$ is given by
$S_A=-\mathrm{Tr}\left(\rho_A\log{\rho_A}\right)$ with the reduced
density matrix defined as $\rho_A=\text{Tr}_B \rho$.
However, entanglement entropy does not
serve as a good entanglement measure for systems with mixed quantum
states as it includes contributions from both classical and quantum
correlations. Such discrepancy leads us to study other entanglement
measures, for instance, mutual information, entanglement negativity
\cite{Vidal:2002zz,Plenio:2005cwa} and the reflected
entropy \cite{Dutta:2019gen}.  We compute in this section the entanglement
entropy, mutual information and the reflected entropy for the massless
Lifshitz theory \eq{az3} for arbitrary $z$.

  \subsection{Entanglement entropy and mutual information}

In this subsection, we apply the formula \eq{Zn} for the trace of the reduced
density matrix to determine the entanglement entropy and mutual information
for bipartite subsystems in the Lifshitz ground state of the theory.
By definition, for a subsystem $A$ in the
   system, the R\'{e}nyi entropy is given by
\begin{equation}
  S_n (A) =\frac{1}{1-n}\log \frac{\Tr(\rho_A^n)}{(\Tr \r_A)^n},
	\label{renyi}
\end{equation}
where $\r_A $ is the reduced density matrix;
and the entanglement entropy can be obtained as
\begin{equation}
	S(A)=\lim_{n\rightarrow1}S_n (A).
\end{equation}
The mutual
information between two subsystems $A$, $B$
is defined by
\begin{equation}	\label{MI_gen}
  I(A:B)=S(A)+S(B)-S(A\cup B).
\end{equation}
By construction, mutual information provides a measure of the correlation
between the two subsystems. If the system is pure, then the mutual information
is twice the entanglement entropy of the state.

\vspace{0.5cm}
\noindent{\bf 1. A finite subsystem in an infinite system: } For a
finite subsystem $A$ of length $l$ in an infinite system,
the trace $\cZ_n$ is given by
\begin{equation}
  \mathcal{Z}_n
  =\cZ^{-n} \int d\phi_1\int d\phi_2 K(\phi_1,\phi_2;l)^n.
			\label{partition}
\end{equation}
Using \eq{K}, we obtain the R\'{e}nyi entropy
\begin{equation}
  S_n(A)= \frac{z-1}{2}\log \frac{l}{\e} +\frac{c_n}{2}	\label{renyi 2}
\end{equation}
and the entanglement entropy
\be \label{ee-inf}
S(A) = \frac{z-1}{2}\log \frac{l}{\e}+  \frac{c_1}{2},
\ee
where  $\epsilon$ is the UV cut-off
and $c_n =  \log\frac{\pi}{\g} + \frac{\log n}{n-1}$ and $c_1 =
\log\frac{\pi}{\g} +1$ are $z$-dependent constants.
Now since $A$ and its complement $B =A^c$ form the total system
which is in a pure state, the mutual information is equal to twice of $S(A)$.

A few remarks follow:
$a)$ We note that the
choice of vacuum does not affect the universal UV part, but only
the finite part of the entanglement entropy through the
constant $\g$.
$b)$ We note that the R\'{e}nyi entropy and the entanglement entropy
share the same universal UV part and differs only in their constant terms. This
is different from the usual case of a conformal vacuum where their UV parts are
proportional with a nontrivial $n$-dependent coefficient:
$[S_n(A)]_{\rm UV} = \frac{1}{2} (1+1/n) [S(A)]_{\rm UV}$. This confirm
that the Lifshitz vacuum is different from the conformal vacuum.

\begin{figure}[H]
			\centering
                        \includegraphics[scale=1.2]{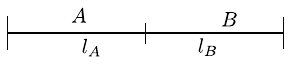}
			\caption{Schematic of two adjacent intervals.}
			\label{Adjacent-config}
		\end{figure}

\vspace{0.5cm}
\noindent{\bf 2. Two adjacent intervals in a finite system:}
Next, we consider the case of a finite length divided
into two adjacent intervals $A$ and $B$ with length $l_A$ and $l_B$
respectively.
Here $A\cup B$ constitutes the
whole system such that the junction point between $A$ and $B$ contains
a free field $\phi$ and the endpoints of the theory have the
Dirichlet boundary conditions satisfying $\phi(0)=0$
and $\phi(l_A+l_B)=0$. See \cref{Adjacent-config}.
In this case, the trace $\cZ_n = \Tr \r_A^n$ is given by
\begin{equation}    \label{partition-adjacent}
  \mathcal{Z}_n
  = \mathcal{Z}^{-n}\int \mathcal{D}\phi~
K(0,\phi;l_A)^n K(\phi,0;l_B)^n=\mathcal{Z}^{-n}\int\mathcal{D}\phi~
e^{-n\g\left(\frac{1}{l_A^{z-1}}+\frac{1}{l_B^{z-1}}\right)\phi^2}.
\end{equation}
Using this,  we obtain the R\'{e}nyi entropy
\begin{equation}    \label{EE_1_DBC}
  S_n(A)=\frac{1}{2}\log{\frac{(l_Al_B)^{z-1}}
    {\left(l_A^{z-1}+l_B^{z-1}\right)\epsilon^{z-1}}}+\frac{1}{2}c_n
\end{equation}
and the entanglement entropy $S_1(A)$.
Note that $S_1(A)$ reduces to \eq{ee-inf} in the case $l_B \to \infty$
of an infinite system.
As above, the mutual information is equal to twice
of $S(A)$.

\begin{figure}[H]
\centering
\includegraphics[scale=1.2]{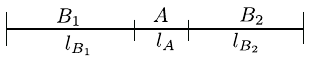}
\caption{Schematic of two disjoint intervals in a finite system.}
			\label{Disjoint-config}
\end{figure}

\vspace{0.5cm}
\noindent{\bf 3. Two disjoint intervals in a finite system:}
Finally we consider the case of a subsystem $A$ of length $l_A$ whose
complement consists of two disjoint intervals
$B_1$ and $B_2$ with lengths $l_{B_1}$ and $l_{B_2}$ respectively.
The geometry is depicted in \cref{Disjoint-config}.
The total length of
the system is $L=l_{B_1}+l_A+l_{B_2}$.
In this case, the trace $\cZ_n$ is given by
\be
\cZ_n = \cZ^{-n} \int \cD \phi_1 \cD
\phi_2 K(0,\phi_1,l_{B_1})^n K(\phi_1, \phi_2,l_A)^n
K(\phi_2, 0, l_{B_2})^n.
\ee
From which, we obtain immediately the R\'{e}nyi entropy as
\begin{equation}\label{EE_no_DBC}
  S_n(A)=\frac{1}{2}\log{\frac{l_{B_1}^{z-1}l_A^{z-1}l_{B_2}^{z-1}}
    {\left(l_{B_1}^{z-1}+l_A^{z-1}+l_{B_2}^{z-1}\right)\epsilon^{2(z-1)}}}+ c_n.
\end{equation}
We note that when $l_{B_1} \to \e$ the system is reduced to a union of two
adjacent intervals as in case 2,
the expression \eq{EE_no_DBC} for $n=1$ reproduces immediately
the length dependence in \eq{EE_1_DBC}. The constant terms also match if
the two UV-cutoffs are chosen to be related appropriately.
		
Next we consider the mutual information between two subsystem $B_1$ and $B_2$
as in figure \ref{Disjoint-config}, whose density matrix is given by
$\r_{B_1 \cup B_2} = \Tr_A \rho$. To determine the entanglement entropy from
this state, we consider the moment of
the reduced matrix $\rho_{B_1} = \Tr_{B_2} \r_{B_1 \cup B_2}$. We obtain
\begin{equation}\label{partition_A_disj}
  \mathcal{Z}_n=\text{Tr}(\rho_{B_1})^n
  =\mathcal{Z}^{-n}\int \mathcal{D}\phi~ K(0,\phi;l_{B_1})^n K(\phi,0;l_{AB_2})^n,
\end{equation}
where $K(\phi,0;l_{AB_2})$ is the propagator
\begin{align}\label{K_A_disj}
  K(\phi, 0;l_{AB_2})&=\int \mathcal{D}\phi'
  K(\phi, \phi';l_A) K(\phi',0;l_{B_2})
\end{align}
is the propagator
over the interval $A\cup B_2$ with a free field
$\phi'$ at the junction of $A$ and $B_2$. Here
$l_{AB_2}^{-(z-1)} = l_A^{-(z-1)} +l_{B_2}^{-(z-1)}$ as was determined in \eq{harm}.
Note that  instead of
$K(\phi, 0;l_A + l_{B_2})$ which might be expected naively,
it is the propagator $ K(\phi, 0;l_{AB_2})$
which appears in \eq{partition_A_disj}.
Using  \eq{K_A_disj}, the trace \eq{partition_A_disj}
can be computed and we obtain
the entanglement entropy for the subsystem $B_1$ where the rest of the
system is partitioned into two subsystems $B_2$ and $A$ as
\begin{equation}\label{EE_A_disj}
  S(B_1)=\frac{1}{2}\log{\frac{l_{B_1}^{z-1}(l_A^{z-1}+l_{B_2}^{z-1})}
    {\left(l_{B_1}^{z-1}+l_A^{z-1}+l_{B_2}^{z-1}\right)\epsilon^{z-1}}}+ \frac{c_1}{2}.
		\end{equation}
Similarly the entanglement entropy for $B_2$ can also be written as follows
\begin{equation}\label{EE_B_disj}
  S(B_2)=\frac{1}{2}\log{\frac{l_{B_2}^{z-1}(l_A^{z-1}+l_{B_1}^{z-1})}
    {\left(l_{B_1}^{z-1}+l_A^{z-1}+l_{B_2}^{z-1}\right)\epsilon^{z-1}}}+ \frac{c_1}{2}.
		\end{equation}
Exploiting the fact that the total system is in a pure state, we get
the expression for $S(B_1\cup B_2)=S(A)$ from \eq{EE_no_DBC}. As a result,
the mutual information between $B_1$ and $B_2$ is given by
\begin{equation}
  I(B_1:B_2)=\frac{1}{2}\log{\frac{\left(l_{B_1}^{z-1}+l_A^{z-1}\right)
      \left(l_{B_2}^{z-1}+l_A^{z-1}\right)}{l_A^{z-1}
      \left(l_{B_1}^{z-1}+l_A^{z-1}+l_{B_2}^{z-1}
      \right)}}=\frac{1}{2}\log{\frac{1}{1-\tilde{\eta}}}~,
			\label{MI_disj}
\end{equation}
where the cross-ratio $\tilde{\eta}(z)$ is given by
\begin{equation}\label{etat}
  \tilde{\eta}(z):=\frac{\left(l_{B_1} l_{B_2}\right)^{z-1}}
        {\left(l_{B_1}^{z-1}+l_A^{z-1}\right)\left(l_{B_2}^{z-1}+l_A^{z-1}\right)}=
        \left[\left(1+\left(\frac{l_A}{l_{B_1}}\right)^{z-1}\right) \left(1+\left(\frac{l_A}{l_{B_2}}\right)^{z-1}\right)\right]^{-1}.
\end{equation}
Notice that all our entanglement measures are entirely expressed in terms of cross ratios
 with components of the form $\sim (\prod_i l_i^{z-1})/(\sum_j l_j^{z-1})$. Moreover,
 from the second expression of \eq{etat}, we immediately observe the importance of
 the ratios of the subsystem lengths in the entanglement observables. When
 $l_A\ll l_{B_i}$, then $\tilde{\eta}(z)\rightarrow 1$. The same happens for
 $l_A< l_{B_i}$ and $z\gg1$. In agreement with expectations the mutual information
 maximizes in these cases since the interactions of the theory have increasing
 range while the length $l_A$ is small compared to the rest subsystems sizes.
 For equal subsystem lengths  $l_A= l_{B_i}$, $\tilde{\eta}(z)= 1/4$, is
 independent of $z$ and as a result the mutual information does not depend
 on $z$ as well.

A further question that naturally arises is whether our theory in the Lifshitz
ground state respects the monotonicity theorems along the RG flow, when the
$c$-function is defined via the entanglement entropy.  The monotonicity of
the $c$-function along the RG flow is guaranteed by the strong subadditivity
in theories with Lorentz symmetry that are unitary. An appropriate $c$-function
candidate for $d$-dimensional anisotropic theories has been suggested
in \cite{Chu:2019uoh} based on previous developments \cite{Casini:2004bw,Ryu:2006ef,Myers:2012ed,Cremonini:2013ipa}. For the 2-dimensional theory it takes the simple usual form
\be\label{c_function}
c=\b \frac{\partial S}{\partial  \ln l}~,
\ee
where $\b$ is a constant of normalization. Applying this definition
of the $c$-function to any of the entanglement formulas in this section,
for example to \eq{ee-inf}, we find that the right monotonicity along the
RG flow, which is for $c'(l)<0$, is guaranteed when $z\ge1$. This constraint
matches nicely the condition of having causality in the theory and satisfies
 maximally the null energy conditions of the holographic dual theory.

\subsection{Reflected entropy and Markov gap}\label{refl-z}

In general, given a mixed state
density matrix $\r$ on a Hilbert space
$\cH_R$ of finite dimension. It is always possible to purify $\rho$ in the sense
that one can always construct a second
Hilbert space $\cH_S$ and a pure state $\ket{\psi}$ such that $\rho$ is given
by the partial trace of $\ket{\psi}\bra{\psi}$ with respect to $\cH_S$.
Now, for a bipartite system in  an arbitrary mixed state
$\rho_{AB}$ on a finite Hilbert space $\mathcal{H}_{A}\otimes\mathcal{H}_{B}$,
there is a canonical purification defined by
a pure state $|\sqrt{\rho_{AB}}\rangle$ in a
doubled Hilbert space $\mathcal{H}_{A}\otimes\mathcal{H}_{B}\otimes
\mathcal{H}_{A^*}\otimes\mathcal{H}_{B^*}$ where $A^*$ and $B^*$ are
dual copies of $A$ and $B$ respectively such that
\be
\rho_{AB} = \Tr_{A^* B^*} (|\sqrt{\rho_{AB}}\rangle \langle \sqrt{\rho_{AB}}|).
\ee
The reflected
entropy is defines as the von Neumann entropy of the reduced density
matrix
$\rho_{AA^*}=\text{Tr}_{BB^*}(|\sqrt{\rho_{AB}}\rangle\langle\sqrt{\rho_{AB}}|)$
\begin{equation}
  S^R(A:B)= -\operatorname{Tr}_{AA^*}\left(\rho_{A A^*}
    \log \rho_{A A^*}\right).
\end{equation}

Markov gap has been proposed
as a  measure of tripartite entanglement \cite{Zou:2020bly}. It
is defined as the
difference between the reflected entropy $S_R(A:B)$ and the mutual
information $I(A:B)$
\begin{equation}\label{MG}
                        h(A:B)=S_R(A:B)-I(A:B).
\end{equation}
As the reflected entropy is lower
bounded by the mutual information \cite{Dutta:2019gen},
Markov gap is non-negative.

The replica method for reflected entropy considers two replica indices
$m$ and $n$, where the former is related to the purification
$|\rho_{AB}^{m/2}\rangle$ of $\rho_{AB}^m$ for positive even integer $m$,
and the latter denotes the usual
R\'{e}nyi index. The reflected density matrix is then given by
\begin{equation}
  \rho_{AA^{*}}^{(m)}=\text{Tr}_{BB^*}\left(
  |\rho_{AB}^{m/2}\rangle\langle \rho_{AB}^{m/2}|
  \right)
\end{equation}
and the $(m,n)$-R\'{e}nyi reflected entropy is defined as,
\begin{equation}\label{S_R_trace}
  S^R_{m,n}(A:B)=\frac{1}{1-n}\log\left[\frac{\text{Tr}
      \left(\rho_{AA^*}^{(m)}\right)^n}{\left(\text{Tr}\rho_{AA^*}^{(m)}\right)^n}
    \right].
\end{equation}
The reflected entropy is obtained by setting $m\rightarrow 1$ followed
by another limit $n\rightarrow 1$.

\vspace{0.5cm}		
\noindent{\bf 1. System of two adjacent intervals:}
Let us first
consider a bipartite configuration of a finite system
with Dirichlet boundary conditions where the two adjacent subsystems $A$ and $B$ have lengths
$l_A$ and $l_B$ respectively as shown in
\cref{Adjacent-config}. It is not difficult to obtain
the trace
$\cZ_{m,n} := \Tr \left(\rho_{AA^*}^{(m)}\right)^n$ as

\begin{equation}\label{Zmn_adj}
\mathcal{Z}_{m,n}
=\mathcal{Z}^{-(m-2)n}\int_{-\infty}^{\infty}\int_{-\infty}^{\infty}d\phi_1
                                d\phi_2 K(0,\phi_1;l_A)^n
                                  K(0,\phi_1;l_B)^n
                                  K(0,\phi_2;l_A)^n
                                  K(0,\phi_2;l_B)^n,
		\end{equation}
where the propagator is given by \eq{K} and $\phi_1$, $\phi_2$
are the fields present at the interface between $A$ and $B$.
Using the expression for the kernel,
we obtain the $(m,n)$-reflected entropy for two adjacent
intervals
\be \label{reflected_adj_mn}
 S^{R}_{m,n}(A:B)=\log\left[
    \frac{(l_Al_B)^{z-1}}{(l_A^{z-1}+l_B^{z-1})\epsilon^{z-1}}
      \right]+c_n
\ee
and the reflected entropy
\begin{equation}
  S^{R}(A:B)=\log\left[
    \frac{(l_Al_B)^{z-1}}{(l_A^{z-1}+l_B^{z-1})\epsilon^{z-1}}
      \right]+c_1.
\label{reflected_adj}
\end{equation}
Note that the $(m,n)$-reflected entropies for the Lifshitz ground states
are independent of $m$, and the dependence on $n$ only appears
in the finite constant piece.
Note also that in general for a  bipartite pure state, the reflected entropy
becomes twice the entanglement entropy. This is exactly what we find here
for \eq{reflected_adj} and \eq{EE_1_DBC}.

As for the Markov gap, it is exactly zero for this specific
configuration since the two adjacent subsystems constituting
the whole system is a pure state. As a result no
tripartite entanglement should be detected from the study of this
bipartite state. Markov gap being zero correctly
serves as a consistency check of our results.

\vspace{0.5cm}		
\noindent{\bf 2. System with two disjoint intervals:}
Next let us consider a bipartite mixed state configuration of
two disjoint intervals on a finite system with Dirichlet boundary conditions in the
boundary. Here we will consider the disjoint subsystems which include
the boundaries of the whole system on an interval although it can be
generalized for any two intervals at arbitrary positions. This
configuration is illustrated in \cref{Disjoint-config}.
We obtain the trace similar to \cite{Berthiere:2023bwn} as
			\begin{equation}
			\begin{split}
				\mathcal{Z}_{m,n}=&\mathcal{Z}^{-(m-2)n}\int
                                d\phi_1d\phi_2....d\phi_{2n}K(0,\phi_1;l_{B_1})^m
                                K(\phi_1,\phi_2;l_A)^{\frac{m}{2}}K(0,\phi_2;l_{B_2})^m
                                K(\phi_2,\phi_3;l_A)^{\frac{m}{2}}\\&\times
                                K(0,\phi_3;l_{B_1})^m
                                .....K(\phi_{2n-1},\phi_{2n};l_A)^{\frac{m}{2}}
                                K(0,\phi_{2n};l_{B_2})^m
                                K(\phi_{2n},\phi_1;l_A)^{\frac{m}{2}},
			\end{split}
		\end{equation}
where $\phi_i$ and $\phi_{i+1}$ are the fields present at the junction
point of ${B_1}\cup {B_2}$ and $A$ and the kernel follows the same form
presented in the calculation for adjacent intervals. Using the kernel
\eq{K}, we obtain
\begin{equation}
  \frac{\mathcal{Z}_{m,n}}{\left(\mathcal{Z}_{m,1}\right)^n}
  =\frac{(\text{det}\mathcal{M}_{m,1})^{\frac{n}{2}}}
  {(\text{det}\mathcal{M}_{m,n})^{\frac{1}{2}}},
\label{partition ratio}
\end{equation}
where $\mathcal{M}_{m,n}$ is the $2n\times 2n$ matrix which takes the same form as \cite{Berthiere:2023bwn}
\begin{equation}
  \mathcal{M}_{m,n}= \frac{m\g}{2l_A^{z-1}}\begin{pmatrix}
    a & -1 & 0 & \ldots & 0    & -1\\
    -1 & b & -1 & 0 & \ldots & 0 \\
    0 & -1 & a  & -1 & \ddots & \vdots\\
    \vdots & \ddots & \ddots & \ddots & \ddots & 0\\
    0 & \ldots & 0 & -1 & a & -1 \\
    -1 & 0 & \ldots & 0 & -1 & b
\end{pmatrix}
\label{matrix 1}
\end{equation}
and $a=2\left(1+\frac{l_A^{z-1}}{l_{B_1}^{z-1}}\right)$,
$b=2\left(1+\frac{l_A^{z-1}}{l_{B_2}^{z-1}}\right)$. The determinant is given by
\begin{equation}	\label{determinant}
  \text{det} \mathcal{M}_{m,n}= \big(\frac{m\g}{2l_A^{z-1}}\big)^{2n}
  4 \sinh^2{n\theta},
  \ee
  where
  \be
 \theta
 =\cosh^{-1}\left(\frac{\sqrt{ab}}{2}\right)
 =\cosh^{-1} \frac{1}{\sqrt{\tilde\eta(z)}}		
\end{equation}
and $\tilde{\eta}(z)$ is the cross ratio \eq{etat}.
As a result, the R\'{e}nyi reflected entropy is
\begin{equation}
  S^R_{m,n}({B_1}:{B_2})=\frac{1}{n-1}\log{\frac{(\sqrt{1-\tilde{\eta}}+1)^{2n}-
      \tilde{\eta}^n}{((\sqrt{1-\tilde{\eta}}+1)^{2}-\tilde{\eta})^n}}.
		\end{equation}
Taking the two consecutive limits
                $m\rightarrow 1$ and $n\rightarrow1$, we obtain the
                expression for the reflected entropy between two
                disjoint intervals as follows,
\begin{equation}
  S^R({B_1}:{B_2})=\frac{1}{\sqrt{1-\tilde{\eta}}}
  \log{\left(\frac{1+\sqrt{1-\tilde{\eta}}}{\sqrt{\tilde{\eta}}}\right)}
  -\log{\left(2\sqrt{\frac{1-\tilde{\eta}}{\tilde{\eta}}}\right)}.
			\label{SR_disj}
\end{equation}

\begin{figure}[H]
\centering
\includegraphics[width=0.5\textwidth]{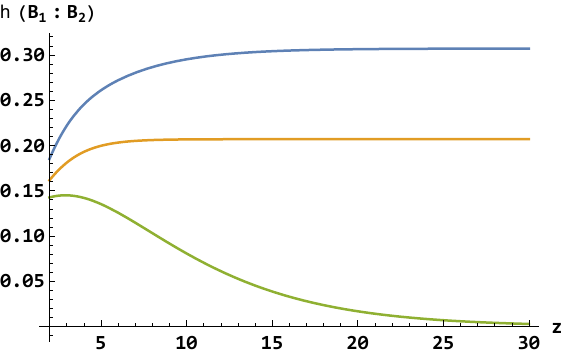}
\caption{The Markov gap as function of $z$ for two disjoint intervals ${B_1}$ and ${B_2}$. It depends on the relative size of the separation interval $A$ in comparison with  the minimum length of ${B_1}$ and ${B_2}$. When $l_A$ is not the minimum length of the three,
 the Markov gap, approaches to zero for large $z$ (green curve), otherwise it saturates to a finite value.}
			\label{MG_1}
\end{figure}

As for the Markov gap, it is
\begin{equation}
  h({B_1}:{B_2})=\frac{1}{\sqrt{1-\tilde{\eta}}}
  \log{\left(\frac{1+\sqrt{1-\tilde{\eta}}}{\sqrt{\tilde{\eta}}}\right)}
  -\log{\left(\frac{2(1-\tilde{\eta})}{\sqrt{\tilde{\eta}}}\right)},
		\end{equation}
where $\tilde{\eta}$ is the cross-ratio. Here we observe a peculiar
characteristics of $\tilde{\eta}$ and the $h({B_1}:{B_2})$ with the variation
of $z$ depending on the sizes of the intervals. For
$l_A<\mathrm{min}\{l_{B_1},l_{B_2}\}$ and $l_A=\mathrm{min}\{l_{B_1},l_{B_2}\}$,
$\tilde{\eta}$ increases and saturates at the values 1 and
$\frac{1}{2}$ respectively. Whereas for $l_A>\mathrm{min}\{l_{B_1},l_{B_2}\}$,
$\tilde{\eta}$ shows a decreasing characteristics followed by an brief
initial increasing phase and finally converges to 0. Similar to the
cross-ratio, the Markov gap also show distinctive behavior depending on
the sizes of the subsystem. In \cref{MG_1}, we have plotted the Markov
gaps with increasing $z$. For $l_A\leq \mathrm{min}\{l_{B_1},l_{B_2}\}$,
$h({B_1}:{B_2})$ increases up to a constant value whereas for $l_A>
\mathrm{min}\{l_{B_1},l_{B_2}\}$, $h({B_1}:{B_2})$ decays to zero. From this
observation, we conclude that with increasing degrees of anisotropy of
the Lifshitz field theory, the tripartite entanglement can be enhanced or completely
destroyed depending on the sizes of the partitions.

\section{Holography}

In this section, we study the holographic dual representation of the Lifshitz
field theory. Lifshitz holography \cite{Kachru:2008yh} has been discussed
extensively in the literature, see for example \cite{Taylor:2015glc} for review.
We will use our field theory result for the entanglement
entropy to fix the $z$-dependence of the radius scale of the bulk Lifshitz
background. And using that, we make a prediction for the time-like entanglement
entropy in Lifshitz field theory.

\subsection{$z$-dependent Lifshitz radius}

The standard form of the (2+1)-dimensions Lifshitz metric with one-direction
anisotropy is given by
\begin{equation}
  ds^2=L^2\left[-\frac{dt^2}{r^{2z}}+\frac{dr^2}{r^2}+\frac{dx^2}{r^2}\right].
 \label{poincare metric 2}
\end{equation}
The above metric is not Lorentz-invariant and supports non-relativistic
Lifshitz translational invariance given by
\begin{equation}
  t\rightarrow\lambda^{z}t,~~x\rightarrow \lambda x,~~r\rightarrow
  \lambda r,
	\label{lss-r}
\end{equation}
that is consistent with the Lifshitz symmetry (\ref{lss})
for the field theory.
 This metric appears as solution of the equations of motion of
 the bulk action given by \cite{Taylor:2008tg}
\begin{equation}
  S=\frac{1}{16\pi G}\int d^{3}x\sqrt{-g}
  \left(R+\frac{z^2+1}{2L^2}-\frac{F^2}{4}-\frac{1}{2}M^2A^2\right),
      \label{action}
\end{equation}
where the usual Einstein's gravity theory is deformed by the inclusion of
a massive vector field given as
\begin{equation}
  A=\sqrt{\frac{2}{z}(z-1)}\frac{L}{r^z}dt~~~ \text{with}~~
  M^2=\frac{z}{L^2}
\end{equation}
that breaks the Lorentz symmetry.
At  constant time, the metric \eq{poincare metric 2} is the same as that for
the (2+1)-dimensional AdS. This would imply the same form of holographic
entanglement entropy for space-like interval. This in fact
takes the same form as what we found
above in \eq{ee-inf} for the free theory.
The agreement is due to the fact that
the form of the UV divergent part of entanglement entropy
is universal. In the case of 2d conformal field theory,
the coefficient of the
entanglement entropy is given by the central charge and its value is protected
from interaction due to a non-renormalization theorem. It is
possible that the scaling symmetry of the Lifshitz theory may do the same
job. Assuming so, then the $z$-dependent prefactor
of \eq{ee-inf} matches with holography if the relation
\be \label{llp}
L = (z-1) l_P
\ee
holds for the radius of curvature $L$ of the Lifshitz geometry.
Here $l_P$ is the Planck length in 3-dimensions.
We remark that previous results in the literature on Lifshitz holography are
independent of this $z$-dependence of the curvature scale $L$.
Here we made use of the entanglement entropy from field theory to propose  the
relation \eq{llp} of parameters.

As we have noted above,
the Lifshitz field theory also admit a Fock vacuum $\ket{0}$
which is different from the RK vacuum $\ket{\Psi_0}$ for any value of $z$.
In particular, the Fock vacuum is defined for $z \geq 1$ while the RK vacuum is defined for $z>1$.
We remark that as $z \to 1$,
the Fock vacuum $\ket{0}$ reduces to the usual conformal vacuum and the
entanglement entropy for an interval $A$ in an infinite system is given by
\be
S(A) = \frac{1}{3} \ln \frac{l}{\e}, \qquad \mbox{for the Fock vacuum at $z=1$}.
\ee
Away from $z=1$, holographic analysis inspired by cMERA
\cite{Nozaki:2012zj,Gentle:2017ywk} gives \cite{He:2017wla}
\be \label{cmera}
S(A) = \frac{z}{3} \ln \frac{l}{z\e}
\ee
for the massless Lifshitz field theory with general $z$.
This is different from \eq{ee-inf} which is obtained from direct computation
for  the Lifshitz vacuum $\ket{\Psi_0}$. This means \eq{cmera} is for a
different vacuum. As \eq{cmera} is continuously connected with the $z=1$ result
for the Fock vacuum, it is probably for the Fock vacuum. 
Our proposal for the holographic relation \eq{llp} is for
the Lifshitz vacuum $\ket{\Psi_0}$ of the field theory, and is different
from the well-known Brown-Henneaux relation \cite{Brown:1986nw}
for the CFT vacuum.
That  $z$-dependence appears here in \eq{llp}
is consistent with the fact the RK vacuum respects
a $z$-dependent Lifshitz symmetry.

\subsection{Time-like entanglement entropy}\label{TEE-lift}
In this section, we determine the time-like entanglement
entropy in a Lifshitz field theory using holography. In the simpler case
of AdS${}_3$/CFT${}_2$, the  time-like entanglement entropy has been obtained
holographically.
By embedding the Poincar\'{e} patch in  the global AdS, the authors in
\cite{Doi:2022iyj,Doi:2023zaf} found that  the
length of time-like geodesic connecting appropriate points in the
global coordinates give rises to an nonvanishing imaginary part to the
time-like entanglement entropy. Since the Lifshitz geometry
is available only in the Poincar\'{e} patch \eq{poincare metric 2} and its global
embedding is not known, it is needed to perform the holographic
computation entirely in the Poincar\'{e} patch. This can indeed be done
straight forwardly as we will show now (with the AdS case covered by $z=1$ in
our computation below).

\begin{figure}[H]
			\centering
                        \includegraphics[scale=1]{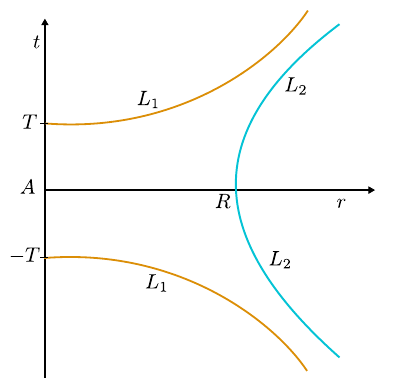}
			\caption{Geodesics for holographic
                          time-like entanglement entropy
                          computed in the Poincar\'{e} patch. The two
                          mirroring curves
that reach the boundary provide the real contribution, while the other bulk
curve provides the imaginary part of the time-like entanglement entropy.}
			\label{TEE-geod}
		\end{figure}

Consider a time interval $A$ in the two dimensional Lifshitz field theory
described by $A\equiv[-T,T]$.  The
geodesics needed for computing the holographic TEE of $A$ consists of
two space-like geodesics connecting the endpoints of $A$ and
infinities plus a time-like geodesic which connects the endpoints of
two space-like geodesics. With the Lifshitz geometry
described in the Poincar\'{e} form
\eq{poincare metric 2}, the equation for the
geodesic on $t-r$ plane is given by
\begin{equation}
t^2=(r^{2z}+c^2)/z^2,
\end{equation}
where $c$ is an integration constant.
Now, depending on value of $c^2$,
we get two classes of geodesics:
	\begin{subequations}
		\begin{align}
		  &t^2=\frac{r^{2z}}{z^2}+T^2, \,\,\,\,\,\, c^2>0
                  \label{geo1}\\&
		  r^{2z}=z^2t^2+R^{2z}, \,\,\,\,\,\, c^2\leq 0,
                  \label{geo2}
			\end{align}
		\end{subequations}
where $T$ and $R$ are constants.
Note that the curve \eq{geo1}  is space-like and describes
a geodesics
extending from the end point $t=T$ (or $t=-T$)
of the time-like interval to infinity. The curve \eq{geo2}
is time-like and, for any value of $R>0$, can be used to smoothly join the
former geodesics at infinity.  The geometry is shown in
figure \ref{TEE-geod}.
Let us denote by $L_1$
the length of the space-like geodesic from the
endpoint $t=T$ to infinity, and by $L_2$ half the
length of the above mentioned connecting time-like geodesic. Then
\begin{equation}
  L_1=LT\int_{\epsilon}^{\infty}\frac{dr}{r}\frac{1}{\sqrt{\frac{r^{2z}}{z^2}+T^2}}
  =\frac{L}{z}\log{\frac{2T}{\epsilon^z}}
		\end{equation}and
		\begin{equation}
 L_2=iLR^z\int_{R}^{\infty}\frac{dr}{r}
                  \frac{1}{\sqrt{r^{2z}-R^{2z}}}=i\frac{L\pi}{2z}.
		\end{equation}
Utilizing the RT formula, the time-like entanglement entropy may be
obtained as
\begin{equation}
\begin{aligned} \label{ST}
S_A^T&=\frac{1}{4 G}\times(2L_1+2L_2)\\
&=\frac{z-1}{2z}\left(\log{\frac{2T}{\epsilon^z}}+\frac{i\pi}{2}\right),
			\end{aligned}
			\end{equation}
where we have used the relation given in \eq{llp}.

We remark that in Lorentz invariant field theory, it has been proposed
to define time-like entanglement entropy in terms of an analytic
continuation based on the Wick rotation.
In the present case, the result \eq{ST} can be obtained by
formally replacing in \eq{ee-inf} the time interval $l$ with the temporal
interval $2T$ as
\be \label{wick}
 l \to (i 2T)^{1/z}.
\ee
The replacement \eq{wick} is consistent with the Lifshitz symmetry
\eq{lss}. However,  due to the presence of fractional derivative, it
is not clear how to implement it on the Lifshitz action as a Wick rotation.
Nevertheless,  the existence of a holographic
time-like entanglement entropy in the Lifshitz case
suggests that there should be a general
definition of time-like entanglement entropy in field theory.
It is interesting to understand better how time like entanglement can be
defined from first principle and to study its physical meaning.

\section{Summary and discussion}\label{summary}

In this paper, we have employed fractional derivative to propose a
definition of the massless Lifshitz theory with arbitrary $z$ in
general $(d+1)$-dimensions. In (1+1)-dimensions, the massless Lifshitz
theory admits a Lifshitz scaling invariant ground state which takes on the form
of RK. Unlike the usual vacuum in
conformal field theory which is defined by the annihilation operators
in the momentum space, the Lifshitz ground state is defined by
annihilation operators in the coordinate space. As a result, we showed
that there is a 2d/1d correspondence between the (1+1)-dimensional Lifshitz
field theory and a dual quantum mechanical system defined with a
fractional derivative. In order for the path integral of the dual
quantum system to be well-defined, a choice of classical solution is
needed to be made. This can be interpreted as that the Lifshitz theory
actually admit a family of ground states whose degeneracy is
parameterized by the choice of the classical solution.

We then computed various bipartite and tripartite entanglement
measures in the Lifshitz theory and determined their $z$-dependence
respectively.
The entanglement measures can be expressed
in terms of simple cross ratios that depend on the lengths of the
intervals and the exponent $z$. Moreover, the standard $c$-function
monotonicity defined via the entanglement entropy along the RG flow,
is satisfied for the range of $z$ that is compatible to causality and
maximally satisfies the null energy conditions. Finally, we considered
a gravity dual
corresponding to the Lifshitz vacuum of the Lifshitz field theory. We showed
that in order to reproduce the field theory result for the
entanglement entropy, the previously considered Lifshitz bulk geometry
has to be supplemented by a Lifshitz radius scale that is dependent on
$z$. Using the dual geometry, we studied and
computed the time-like entanglement entropy for a time-like subsystem.
In Lorentzian theory, the time-like entanglement entropy can be
obtained from the space-like one via an analytic continuation. In
Lifshitz theory, space and time are different due
to their different scaling, and a standard form of Wick rotation
that mixes space and time cannot be introduced.
Our holographic result suggests that
there should be a field theoretic definition of time-like entanglement
entropy that does not employ analytic continuation.
It is an interesting question for further investigation.

The entanglement observables in the massless Lifshitz scalar theory
have been computed in this paper using field theory methods. For the
entanglement entropy of an interval in an infinite system, the result
can be readily reproduced from holography provided that the mapping
relation \eq{llp} is implemented. The situation is more complicated
for entanglement observables defined in a finite system. For CFT, the
dual geometry involves a global AdS, suggesting that the global
completion of the Poincar\'{e} patch of the Lifshitz geometry
\eq{poincare metric 2} for $z\neq 1$ is needed. This is however quite
a non-trivial open problem.  We will leave the holographic
entanglement computation for massless Lifshitz theory on finite system
for future analysis.

There are many interesting directions to follow from our work in this
paper.
Here, we were confined to computing various entanglement
measures in the zero temperature case. It will be interesting to extend our
analysis to non-zero temperature in the context of Lifshitz holography
which still remains a non-trivial issue. The study of time evolution
of entanglement measures in Lifshitz field theory
as a function of $z$ is also interesting.
Our work is expected to bring new insights in the studies of
quantum critical phenomenon in condensed matter systems exhibiting
Lifshitz scaling. It is interesting to consider boundary Lifshitz
field theory (BLFT) and its holography by generalizing  previous AdS/BCFT
formulations with Neumann boundary condition
\cite{Takayanagi:2011zk,Fujita:2011fp}, conformal boundary condition
and Dirichlet boundary conditions \cite{Miao:2017gyt, Chu:2017aab,
  Miao:2018qkc,Chu:2021mvq}. It is also interesting to analyze the
phase structure of time-like entanglement entropy in the
aforementioned BLFT similar to \cite{Chu:2023zah}.  We hope to return
to these exciting issues in the near future.

\section*{Acknowledgement}
JKB and DG acknowledge the support of the National Science and
Technology Council (NSTC) of Taiwan with the Young Scholar Columbus
Fellowship grant 112-2636-M-110-006.
AC and CSC acknowledge support of this work by
NCTS and the grant 110-2112-M-007-015-MY3 of NSTC. HP acknowledges the
support of this work by NCTS.

\begin{appendices}
  \section{Fractional derivative via Fourier analysis}\label{Frac-deriv}
  
\begin{figure}[H]
	\centering
        \includegraphics[scale=0.4]{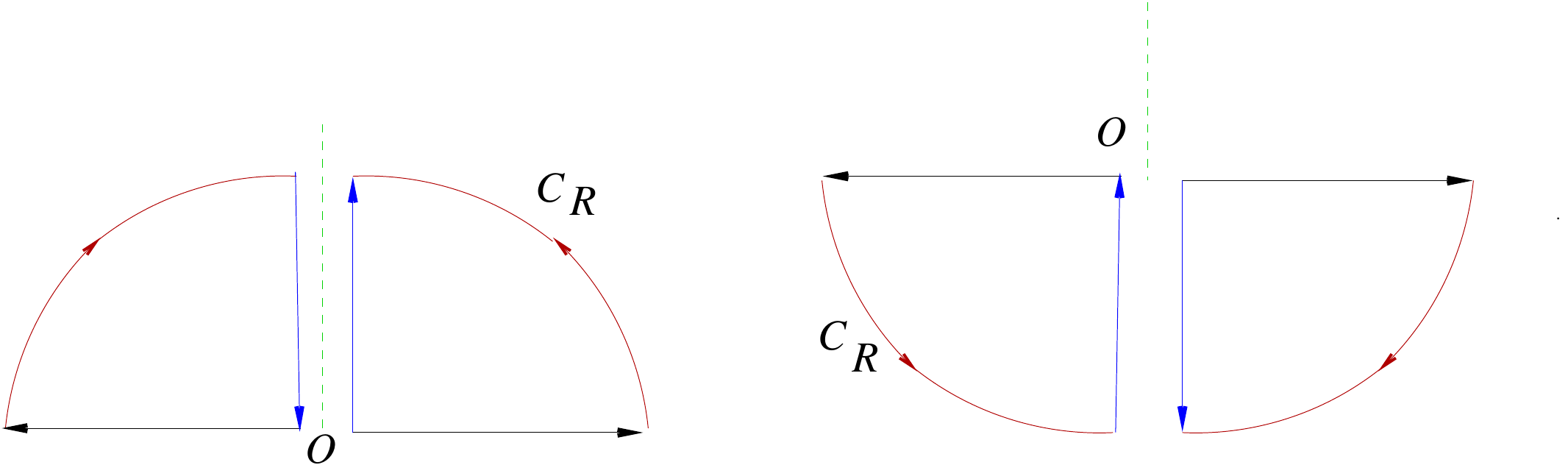}
	\caption{Contour deformation for $x>0$ (left)
          and $x<0$ (right). The line integral on $C_R$ vanishes for $\b>0$
        as $R \to \infty$. The dotted green line denotes the branch cut for $w^z$, $w \in\mathbb{C}$.}
	\label{cont-def}
  \end{figure}

In this appendix, we demonstrate how to determine the fractional derivative of an arbitrary function using (\ref{del-def}). Consider a function that can be expressed in the following form, $\phi(x)  =
\int_C dq \; a(q) e^{i q x}$ for some contour $C$ and some coefficient
$a(q)$. It is important to note that
for arbitrary $\b$ and $z$, the integrand of $\nabla^z_x \phi$
has a branch point singularity at the origin $q=0$  and so
one has to be careful in finding the right choice of
$C$ so that a
solution $\phi$ can be obtained.
A naive choice of $C$ over the real axis would not work.
With an appropriate choice of contour, one can show that
\be\label{del-x}
\del_x^z x^\b = \frac{\G(\b+1)}{\G(\b-z+1)} x^{\b-z} \quad
\mbox{for $x>0$ and for any real $\b, z$}.
\ee
To see this, let us consider
the function defined by
\be \label{Hb}
H_\b (x) := \frac{1}{2\pi}
\int_C
 (\frac{1}{iq})^{\b+1} e^{i q x} dq \qquad \mbox{for $\b>0$},
\ee
where $C$ is the countor $-\infty <q< -\e, \e < q < \infty$
with $\e \to 0$.
It follows immediately from the definition \eq{Hb} that
\be \label{del-Hb}
\nabla_x^z H_\b(x) = H_{\b -z}(x) \quad \mbox{for $\b>0$, $\b>z$}. 
\ee
Now, it is easy to perform a contour deformation and
evaluate the line integral in \eq{Hb}, see figure \ref{cont-def}.
Taking into account of the branch cut, we obtain immediately that
for $\b>0$,
\be
H_\b(x) = \begin{cases}
  \frac{x^\b}{\G(\b+1)} , & x\neq 0 \\
  0, & x<0.
  \end{cases}
\ee
Using this, one can establish \eq{del-x} by noticing that for any $\b, z$, one
can always find a $\b_0 >0$ and $\b_0>z$ such that $\b = \b_0 - n$ for some
positive integer $n$ and so \eq{del-x} holds for $\del_x^z x^{\b_0}$.
Now, differentiate this with the ordinary derivative $\del_x^n$ and use the
commutativity of the derivatives $\del_x, \del_x^z$, we
arrive at \eq{del-x} for any $\b,z$.

	\end{appendices}


\bibliographystyle{JHEP}

\bibliography{Lifshitz}

\end{document}